\documentclass[sigconf]{acmart}

\AtBeginDocument{%
  \providecommand\BibTeX{{%
    \normalfont B\kern-0.5em{\scshape i\kern-0.25em b}\kern-0.8em\TeX}}}

\setcopyright{acmlicensed}
\copyrightyear{2018}
\acmYear{2018}
\acmDOI{XXXXXXX.XXXXXXX}

\acmConference[Conference acronym 'XX]{Make sure to enter the correct
  conference title from your rights confirmation emai}{June 03--05,
  2018}{Woodstock, NY}
\acmISBN{978-1-4503-XXXX-X/18/06}

\usepackage{enumitem} 

\copyrightyear{2024}
\acmYear{2024}
\setcopyright{rightsretained}
\acmConference[AVI 2024]{International Conference on Advanced Visual Interfaces 2024}{June 3--7, 2024}{Arenzano, Genoa, Italy}
\acmBooktitle{International Conference on Advanced Visual Interfaces 2024 (AVI 2024), June 3--7, 2024, Arenzano, Genoa, Italy}\acmDOI{10.1145/3656650.3656666}
\acmISBN{979-8-4007-1764-2/24/06}

\begin{document}

\title[A Design Elicitation Study of How We Visually Express Data Relationships]{What We Augment When We Augment Visualizations: A Design Elicitation Study of How We Visually Express Data Relationships}

\author{Grace Guo}
\orcid{0000-0001-8733-6268}
\affiliation{%
  \institution{Georgia Institute of Technology}
  \streetaddress{North Ave NW}
  \city{Atlanta}
  \state{Georgia}
  \country{USA}
  \postcode{30332}
}
\email{gguo31@gatech.edu}

\author{John Stasko}
\orcid{0000-0003-4129-7659}
\affiliation{%
  \institution{Georgia Institute of Technology}
  \streetaddress{North Ave NW}
  \city{Atlanta}
  \state{Georgia}
  \country{USA}
  \postcode{30332}
}
\email{john.stasko@cc.gatech.edu}

\author{Alex Endert}
\orcid{0000-0002-6914-610X}
\affiliation{%
  \institution{Georgia Institute of Technology}
  \streetaddress{North Ave NW}
  \city{Atlanta}
  \state{Georgia}
  \country{USA}
  \postcode{30332}
}
\email{endert@gatech.edu}


\begin{abstract}
Visual augmentations are commonly added to charts and graphs in order to convey richer and more nuanced information about relationships in the data.
However, many design spaces proposed for categorizing augmentations were defined in a top-down manner, based on expert heuristics or from surveys of published visualizations.
Less well understood are user preferences and intuitions when designing augmentations.
In this paper, we address the gap by conducting a design elicitation study, where study participants were asked to draw the different ways they would visually express the meaning of ten different prompts.
We obtained 364 drawings from the study, and identified the emergent categories of augmentations used by participants.
The contributions of this paper are: (i) a user-defined design space of visualization augmentations, (ii) a repository of hand drawn augmentations made by study participants, and (iii) a discussion of insights into participant considerations, and connections between our study and existing design guidelines.
\end{abstract}




\received{20 February 2007}
\received[revised]{12 March 2009}
\received[accepted]{5 June 2009}

\maketitle

\section{Introduction}
When designing visualizations and visualization systems, it is common for charts and graphs to be augmented with visual elements in order to convey richer and more nuanced information about relationships in the data.
These visual augmentations range from textual and graphical annotations to embellishments and free-form drawings.
They have been used to support different user tasks and usage scenarios from communicating insights and supporting collaboration \cite{heer2008generalized, badam2021integrating, viegas2007manyeyes, ellis2004collaborative} to guiding readers through narrative stories \cite{kosara2013storytelling, lee2015more, hullman2013contextifier}, tracking analytic provenance \cite{ragan2015characterizing, freire2008provenance}, and visualizing implicit error \cite{mccurdy2018framework}.
Augmentations and embellishments have been found to enhance comprehension and recall of visualizations \cite{bateman2010useful, few2011chartjunk, li2014chart, inbar2007minimalism, hullman2011benefitting, borkin2013makes, borgo2012empirical}, while free-form drawings can also reflect playful commentary \cite{heer2007voyagers}, user hunches and ambiguity \cite{lin2021data}.

The effectiveness and broad applicability of visual augmentations have motivated the creation of design spaces to guide how augmentations should be added to visualization systems and tools \cite{ren2017chartaccent, srinivasan2018augmenting}, and to serve as references for designers creating data visualizations \cite{chen2022vizbelle}.
However, many of these design spaces were defined in a top-down manner, based on expert heuristics \cite{srinivasan2018augmenting} or from surveys of published visualizations \cite{ren2017chartaccent, chen2022vizbelle}.
Less well understood are the types of augmentations used when people are given the flexibility of free-form sketching to express various data relationships.
Defining such a design space from the bottom-up can help identify common categories of augmentations that may be more intuitive or interpretable to visualization audiences.
It can also surface examples of augmentation designs that are not well supported by current visualization tools, and provide guidance on how these tools might be extended to better match user expectations and mental models.

We address this gap by conducting a design elicitation study to map out a user-defined visualization augmentation design space.
In prior works, augmentations have been variously referred to as embellishments, chart-junk, annotations, and free-from drawings.
These categories overlap with one another, and we use the term \textit{augmentations} in this paper to refer to them collectively.
Synthesizing from prior work, we define visualization augmentations as \textit{any visual alteration to a chart or graph designed to convey information about data relationships not already captured in the data mappings.}

In our design elicitation study, we provided participants with 10 prompts about different data relationships.
Participants were then asked to draw on a set of printed bar, point, and line charts the different ways they would visually express the meaning of each prompt.
We obtained 364 drawings from the study, and using an affinity diagramming process, identified eight main categories of augmentations used by participants.
Based on our study results, we discuss insights into participant considerations when designing visual augmentations, and highlight connections between our findings and existing visualization design guidelines.
In sum, the contributions of this paper are: (i) a user-defined design space of visualization augmentations used to express data relationships, (ii) a repository of hand drawn augmentations made by study participants, and (iii) a discussion of insights into participant considerations, and connections between our study and existing design guidelines.

\section{Related Work} \label{related-work}

\subsection{Augmenting Visualizations}

There exists a wide variety of visual augmentations that have been explored in prior works, ranging from embellishments and free-form drawings, to textual and graphical annotations.
Perhaps most well known are `chart junk' or visual embellishments, referring broadly to non-essential imagery or ornamentation added to a visualization.
While chart junk was discouraged in many early visualization guidelines \cite{wainer1984display, tufte1985visual}, they have since been found to potentially enhance comprehension and recall \cite{bateman2010useful, few2011chartjunk, li2014chart, inbar2007minimalism, hullman2011benefitting, borkin2013makes, borgo2012empirical}.
More commonly, textual and graphical annotations have been used to visually convey information not already captured in the data.
They help convey computer generated statistics and facts \cite{heer2008generalized, badam2021integrating, viegas2007manyeyes, ellis2004collaborative, ren2017chartaccent, chen2013manyinsights}, guide users through narratives and data stories \cite{kosara2013storytelling, lee2015more, hullman2013contextifier}, track analytic provenance \cite{ragan2015characterizing, freire2008provenance}, and visualize implicit error \cite{mccurdy2018framework}.
Yet still other visual augmentations use free-form drawings to provide playful commentary \cite{heer2007voyagers} and express hunches about data \cite{lin2021data}.
Many of these studies also implemented tools that help users explore and add augmentations to their visualizations.
Some of these, such as \cite{chen2013manyinsights, ren2017chartaccent, heer2007voyagers, badam2021integrating}, leverage an interactive graphical interface, while others are programming toolkits \cite{LuAnnotation, ggAnnotation} that provide developers fine-grained control over the position and appearance of augmentations added to their charts.

The goal of this paper is not to provide definitions for the different types of visual augmentations that have appeared in prior work.
The above examples are not distinct categories, and many of them overlap with one another.
For instance, the terms chart junk and embellishments have been used interchangeably by some researchers \cite{bateman2010useful, li2014chart, borgo2012empirical} but not others.
Similarly, specific augmentations -- such as threshold and regression lines -- have been considered an embellishment in some studies \cite{srinivasan2018augmenting} and an annotation in others \cite{ren2017chartaccent}.
This paper focuses instead on the design space of visual augmentations, including the specific marks used, the types of graphical elements added to charts, and the underlying data relationships conveyed.

\subsection{Augmentation Design Spaces} \label{related-work-design}
In prior studies, a number of design spaces have been proposed that help guide the implementation of visual augmentation tools \cite{ren2017chartaccent, srinivasan2018augmenting}, and to serve as references for amateur designers creating data visualizations \cite{chen2022vizbelle}.
Taken together, these studies demonstrate that design spaces are useful for guiding how users augment visualizations, and how tools can be implemented to support the process.
However, existing design spaces were all defined in a top-down manner, based on expert heuristics \cite{srinivasan2018augmenting} or from surveys of published visualizations \cite{chen2022vizbelle, ren2017chartaccent}.
In our paper, we supplement existing work by conducting a design elicitation study where participants were asked to draw onto printed visualizations how they would visually convey a set of prompts.

\section{Design Elicitation Study}

We adopt the design elicitation methodology to study how visual augmentations are used to convey data relationships in charts and graphs.
We chose this approach in order to 1) define a bottom-up, user-generated design space of visual augmentations, and 2) understand participant considerations and expectations when augmenting visualizations.
In the study, participants were provided with 10 prompts and asked to draw on printed charts how they would use augmentations to visually convey each prompt.

\subsection{Data Set}

We used a fictitious data set of monthly store performance for five stores in a franchise.
The data set had 60 rows, and included attributes: \textit{store name}, \textit{month}, \textit{revenue (in thousands)}, \textit{cost (in thousands)}, and \textit{profit (in thousands)}.
This data set was created to include a variety of categorical, quantitative, and temporal data variables.

\begin{table}[htb]
\includegraphics[width=\columnwidth]{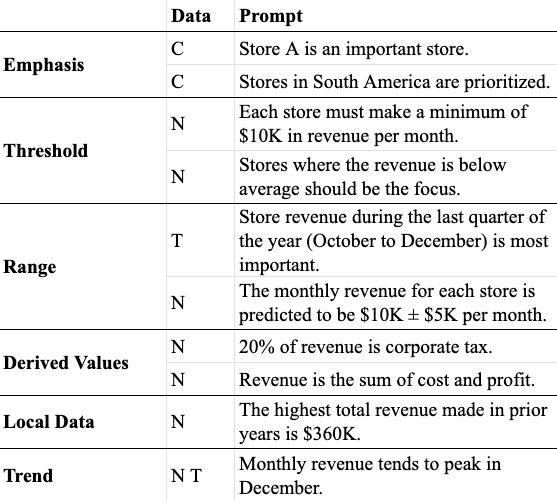}
\caption{\label{tab:prompts}All prompts used in the study (\textbf{N}: Numeric, \textbf{C}: Categorical, \textbf{T}: Temporal).}
\end{table}

\subsection{Prompts} \label{prompts}

From prior works \cite{srinivasan2018augmenting, ren2017chartaccent, chen2022vizbelle}, we first synthesized the most common data relationships included in various augmentation design spaces.
Our synthesis focused on the visual alterations proposed and the data needed to render each augmentation.
For example, stroke (or quadrant) lines were used in both VizBelle and Voder for tasks ranging from emphasizing outliers \cite{chen2022vizbelle} to summarizing distributions \cite{chen2022vizbelle, srinivasan2018augmenting}.
In our synthesis, we consider all of the above as examples of ``threshold'' data relationships that express how each data instance relates to the value indicated by the stroke/quadrant line(s).
In total, we identified six data relationships: \textit{emphasis}, \textit{thresholds}, \textit{ranges}, \textit{derived values}, \textit{local/user-defined data}, and \textit{trends}. 
We then created 10 prompts based on these data relationships (Table \ref{tab:prompts}).
All prompts were iteratively revised based on feedback from 4 rounds of pilot studies to ensure a diversity of framings and data types.

\subsubsection{Emphasis\nopunct}
relationships reflect data values of significance in a particular usage context.
To indicate their importance in a visualization, visual elements corresponding to these values are often highlighted or made more salient \cite{srinivasan2018augmenting, ren2017chartaccent, chen2022vizbelle}.
Emphases can be defined for numerical, categorical, and temporal data variables.

\subsubsection{Thresholds\nopunct}
are data relationships that represent a value at a boundary or transition point \cite{ren2017chartaccent, chen2022vizbelle}.
Threshold values can be defined for numerical and temporal variables, and can be both user-specified or statistically computed.

\subsubsection{Ranges\nopunct} are defined by minimum and maximum value extents \cite{srinivasan2018augmenting, ren2017chartaccent, chen2022vizbelle}.
These values can be numerical or temporal, and can be both user-specified or statistically computed.
The interquartile range, for example, is a common statistic where the minimum (Q1) and maximum (Q3) values define the middle 50\% of data \cite{srinivasan2018augmenting}.

\subsubsection{Derived values\nopunct} are calculated from variables in the data set \cite{srinivasan2018augmenting} and describe how they relate to each other mathematically.
Derived values can be computed from existing numerical variables, and should be specified by users based on their domain knowledge.

\subsubsection{Local/User-defined Data\nopunct} are additional data provided by a user to supplement an existing data set.
These can be determined by prior knowledge, domain expertise, or personal approximations (i.e. ``data hunches'' \cite{lin2021data}).
These supplemental data values should have the same variable types as the existing data set.

\subsubsection{Trends\nopunct} describe how one variable in the data set changes in relation to another variable, and are included in nearly all prior design spaces \cite{srinivasan2018augmenting, ren2017chartaccent, chen2022vizbelle}.
Typically, one of these variables is temporal, while the other varies over time.
Trends can also refer to statistical trends between two numerical variables (e.g. regression, normal distributions etc.).
However, we chose to exclude prompts about statistical trends since prior works \cite{yang2018correlation, breheny2017visualization, correll2017regression, wang2017line} have extensively explored how they can be visually conveyed on charts and graphs.

\subsection{Participants}
We recruited 12 participants through mailing lists and by word of mouth.
10 participants were aged between 23 and 27 (2 preferred not to say).
All participants completed a pre-study questionnaire to assess their ability to read the types of visualizations that would be used in the study.
We planned to exclude respondents if they answered more than 2 questions incorrectly or if their self-reported experience was low, but all respondents satisfied these conditions.
Recruitment continued until empirical saturation, when no new augmentations were observed in study responses.
The study was conducted in person, and took about an hour each.
All participants were compensated with a \$15 gift certificate.

\subsection{Task}

Participants were provided with 10 prompts (Table \ref{tab:prompts}) in randomized order to mitigate order effects.
Each prompt came with 2 point charts, 2 bar charts, and 2 line charts printed on paper.
These chart types were chosen as they were the most commonly used visualization types surveyed in prior works \cite{srinivasan2018augmenting, ren2017chartaccent, chen2022vizbelle}.
Participants were asked to draw on the provided visualizations the augmentations they would use to express the content of the prompt, using \textit{at least one} of each chart type per prompt.
They were allowed to skip chart types, but only in cases where they could provide a justification for why the chart type was unsuitable for the prompt.

In order to encourage ideation and avoid the trivial solution of just copying the provided prompt, we asked participants to avoid text-only augmentations.
However, participants could use text in labels or legends in combination with a visual representation.
All participants were told that they would not be evaluated on drawing precision, and were asked to think aloud while completing the task.







\section{Results} \label{results}
We obtained a total of 364 drawings\footnote{Note that because participants were allowed to skip charts (with justification) or provide additional drawings, the total responses received was not exactly $12 \times 10 \times 3$.}, and used an affinity diagramming process to identify eight categories of augmentations used by participants (Table \ref{fig:categories_count}).
The most common categories were: encodings, marks, threshold lines, segmentation, and shaded ranges.
All participant drawings can be found in our supplemental materials\footnote{https://github.com/auteur-vis/design\_elicitation\_study}.

Our results demonstrate broad consensus with prior works in visualization augmentation.
Encoding channels and threshold lines, for example, have been included in all prior augmentation design spaces \cite{ren2017chartaccent, srinivasan2018augmenting, chen2022vizbelle}, and were implemented in visualization augmentation tools such as Voder \cite{srinivasan2018augmenting}, ChartAccent \cite{ren2017chartaccent}, and ManyInsights \cite{chen2013manyinsights}.
In the following sections, we focus our discussion on unexpected or novel observations from our study results.

\subsection{Encodings} \label{adding-channels}

\begin{figure}[H]
 \centering
 \includegraphics[width=\columnwidth]{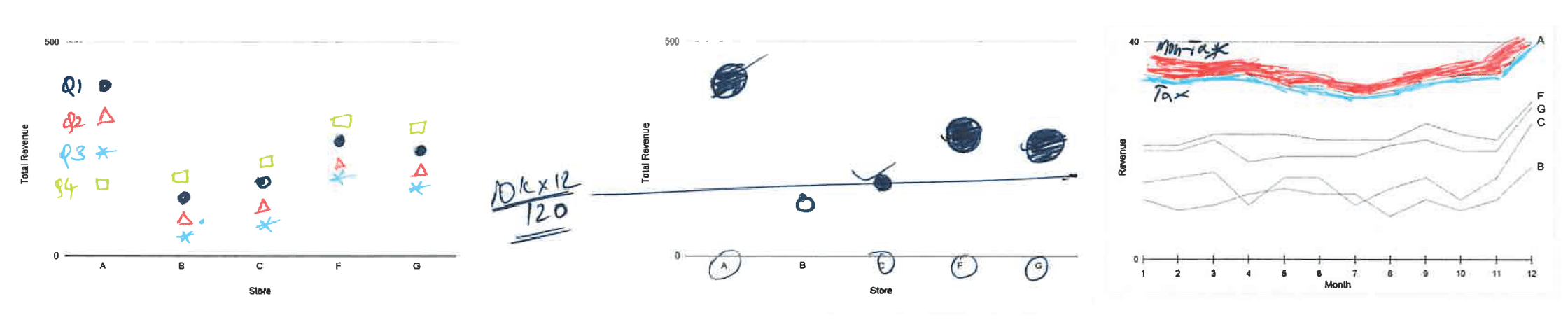}
 \caption{Examples of encoding channels used in the study.}
 \label{fig:encoding_examples}
\end{figure}

The most common augmentation by participants was the addition of new encoding channels.
While color was frequently chosen, study participants also explored using the \textit{opacity}, \textit{size/radius}, \textit{stroke width}, and \textit{shape} encoding channels (Fig. \ref{fig:encoding_examples}).
These other encodings appeared infrequently, possibly because they were applicable to only some of the chart types.
For example, size/radius and shape encodings could only be added to point charts, while stroke width was only added to line charts.
In the case of the opacity encoding, the difficulty of ``drawing'' opacity may have contributed to its infrequent use.

\subsection{Marks} \label{marks}
In addition to encodings, participants also frequently altered the marks used in a chart.
We identified three main categories of such augmentations: adding or layering marks onto a visualization (Fig. \ref{fig:add_mark_examples}), changing the mark (i.e. changing the chart type, for example, from point to bar chart, Fig. \ref{fig:change_mark_examples}), and adding glyphs (Fig. \ref{fig:glyph_examples}).
Of these, glyphs appeared most infrequently, in only two instances.
However, this could be attributed in part to their complexity, making them difficult to draw on paper.
They may also simply be unsuitable for the prompts in our study.
Further work may be necessary to better understand why glyphs were so rarely used, and what implications this has for how visualization systems and tools are designed.



\subsection{Threshold Lines and Shaded Regions} \label{threshold}

Other commonly used visual augmentations were threshold lines (Fig. \ref{fig:threshold_line_examples}) and shaded regions (Fig. \ref{fig:shaded_range}).
Interestingly, we found that participants largely preferred using threshold lines over shaded ranges.
Even for prompts designed to elicit responses about range data relationships, threshold lines were used more in all chart types.

\subsection{Segmentation} \label{segmentation}

\begin{figure}[H]
 \centering
 \includegraphics[width=\columnwidth]{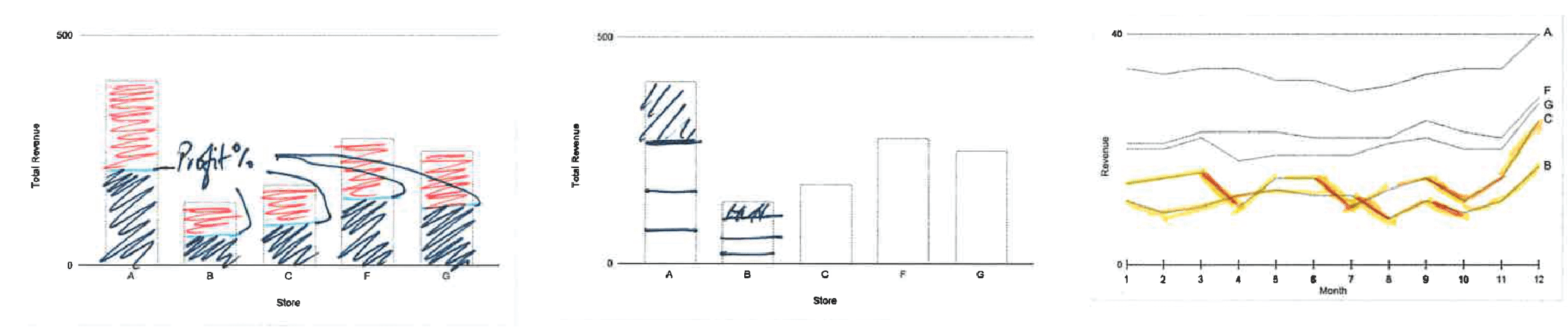}
 \caption{Segmentation applied to bar charts and line charts.
}
 \label{fig:segments_examples}
\end{figure}

Segmentation was often used in bar charts and line charts to divide marks into smaller sub-components to convey information about individual parts that make up a whole.
One visual augmentation strategy that often appeared in concurrence with segmentation was the use of partial fills (Fig. \ref{fig:segments_examples}, middle).
A partial fill occurs when a mark is divided into smaller segments, but a color encoding is applied to only some of the sub-segments.
This was often used to highlight the most relevant subgroups for comparison.

\subsection{Additional Augmentations}
Finally, we discuss augmentations that appeared infrequently in the study, but nonetheless present interesting design alternatives.

\subsubsection{Categorical Scale Change} \label{order}

Although most participants emphasized data values using encodings, participant P11 proposed using order as an indicator of priority (Fig. \ref{fig:order_examples}), explaining that \textit{``I don’t want to change the color because they’re not more important, I just want [viewers] to see it first.''}
Although this approach only applies to categorical variables, it presents an interesting alternative to encoding augmentations that may be useful in other contexts.



\subsubsection{Free Form Drawing} \label{free-form}

Surprisingly, few participants in our study used free form drawings for augmentation (Fig. \ref{fig:free_form_examples}).
These drawings included adding check marks or circling interesting parts of the visualization, and were similar to the informal graphical marks seen in prior work such as Sense.us \cite{heer2007voyagers}.

\section{Discussion} \label{discussion}

While the results of our study broadly agree with many prior works in visualization augmentation, it also surfaced some unexpected patterns in augmentation usage.
In the following sections, we discuss insights into participant considerations, and draw connections between our study and existing design guidelines.

\subsection{User Preferences and Design Best Practices}
In our design elicitation study, color encodings were frequently used across all chart types due to factors of familiarity, attention, and personal associations.
However, color is also strongly influenced by the human perceptual system \cite{silva2011using}, as well as contextual factors and personal associations such as emotion \cite{bartram2017affective}, culture \cite{schirillo2001tutorial} and language \cite{saunders2002trajectory}.
Taken together, this suggests that user preferences for how augmentations (particularly color encodings) should be used may not always agree with visualization design best practices.
In future work, researchers developing augmentation tools may thus consider including recommendation features or system defaults that guide users towards more effective and accessible augmentation choices.

\subsection{Biases and Misinformation}

Additionally, we also want to consider how augmentations can introduce biases and misinformation into visualizations, and the implications of this outcome.
Cognitive biases often affect how visualizations are read and interpreted \cite{wall2019formative, dimara2018task}, and may be exacerbated by augmentations.
Threshold lines (Section \ref{threshold}), for example, might create an anchoring bias where readers overly focus on data around the reference value.
Similarly, the frequent use of color encodings may cause readers to overemphasize the highlighted chart elements.
Augmentations can also introduce misinformation into a visualization.
User generated augmentations are subject to human errors -- intentional and unintentional.
By adding them to a visualization, these errors may then be amplified and legitimized \cite{karduni2018can, boyd2021can, lauer2020deceptive}.
In prior work looking at data hunches \cite{lin2021data}, Lin et al. recommended distinguishing augmentations from the original visualization by giving them a ``sketchier'' appearance.
This helps to visually indicate the uncertainty associated with user-provided data, and may be useful in combating the effects of misinformation.
Future work can further investigate the impact of augmentations on bias and misinformation, as well as other strategies to mitigate their effects.

\subsection{Limitations}

Our study was limited in terms of the chart types included and statements used as prompts, and further work is needed to understand how visual augmentations might be designed for other chart types.
Additionally, while we aimed to include prompts that would elicit a broad range of augmentations from our participants, it is possible that there are statements specific to certain usage scenarios not considered in our study.
Finally, the results of our study may also be constrained by pen and paper hand-drawing methods.
For example, the convex hulls in Voder \cite{srinivasan2018augmenting} and ChartAccent \cite{ren2017chartaccent}, as well as the use of images and clip-art in ChartAccent \cite{ren2017chartaccent} and VizBelle \cite{chen2022vizbelle}, may have all been under-represented as they are difficult to draw by hand.

\section{Conclusion}
In this paper, we conducted a design elicitation study to understand and define a space of visualization augmentation designs from a bottom-up, user-centered approach.
We obtained 364 drawings, from which eight main categories of visual augmentations were identified.
For each category, we provided definitions, examples, as well as insights into participant processes and considerations.
Finally, we highlighted connections to existing visualization design spaces and guidelines.
All visual augmentations drawn by participants have been open-sourced for further analyses.
In future work, we see potential for synthesizing existing design spaces and the results of this work into a single, comprehensive reference to guide the design and implementation of tools to support visualization augmentation.

\begin{acks}
The authors would like to thank the Georgia Tech Visualization Lab for their feedback and suggestions. This work is supported in part by NSF IIS-1750474 and the IBM PhD Fellowship.
\end{acks}

\bibliographystyle{ACM-Reference-Format}
\bibliography{sample-base}

\appendix

\section{Augmentation Category Counts}
The raw counts of various augmentation categories observed in our design elicitation study are reported in Table \ref{fig:categories_count}.

\begin{table*}[h]
 \centering
 \includegraphics[width=\textwidth]{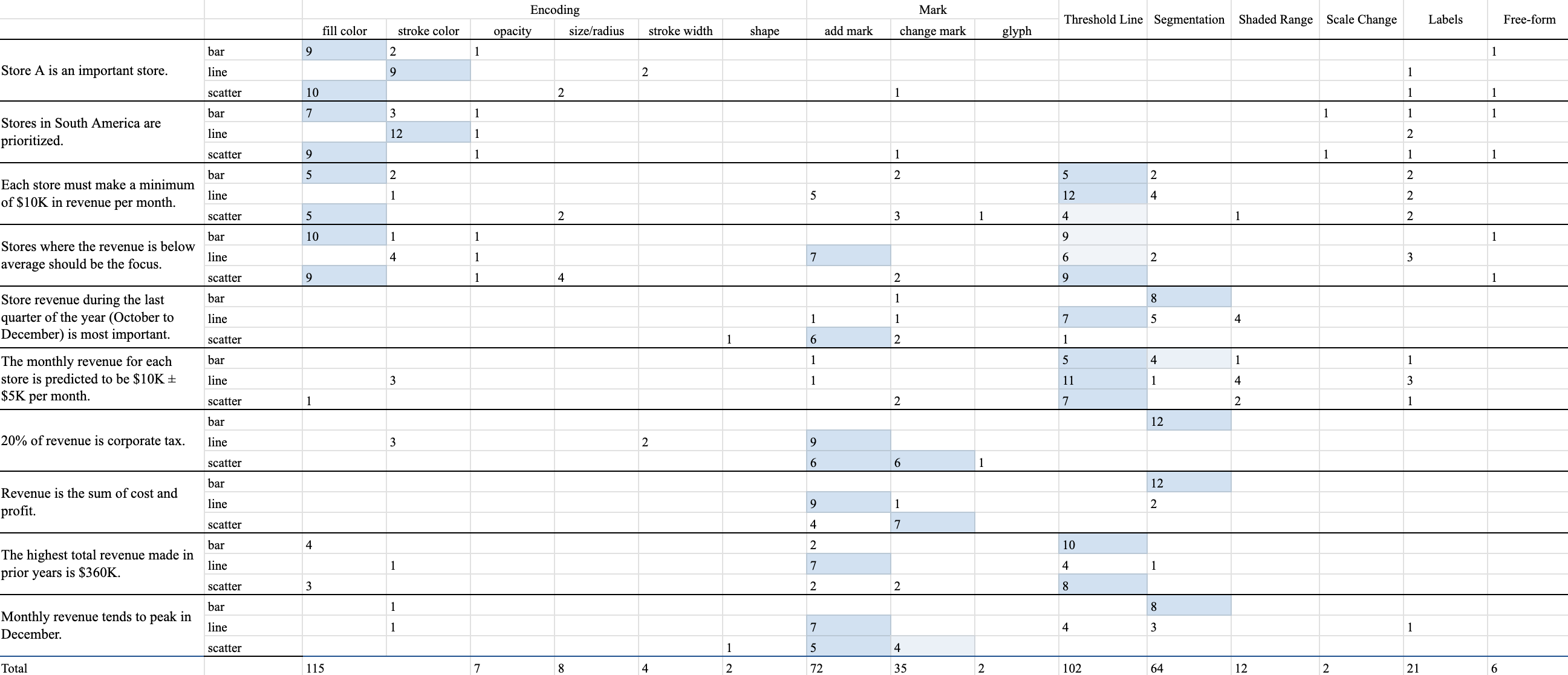}
 \caption{How frequently each augmentation category was used by study participants for a given combination of prompt and chart type. If more than one augmentation was used in the same visualization, we counted the visualization twice. The most common augmentations are highlighted in blue. The calculated totals of fill color and stroke color have been combined.}
 \label{fig:categories_count}
\end{table*}

\section{Participant Drawings}

\begin{figure}[H]
 \centering
 \includegraphics[width=\columnwidth]{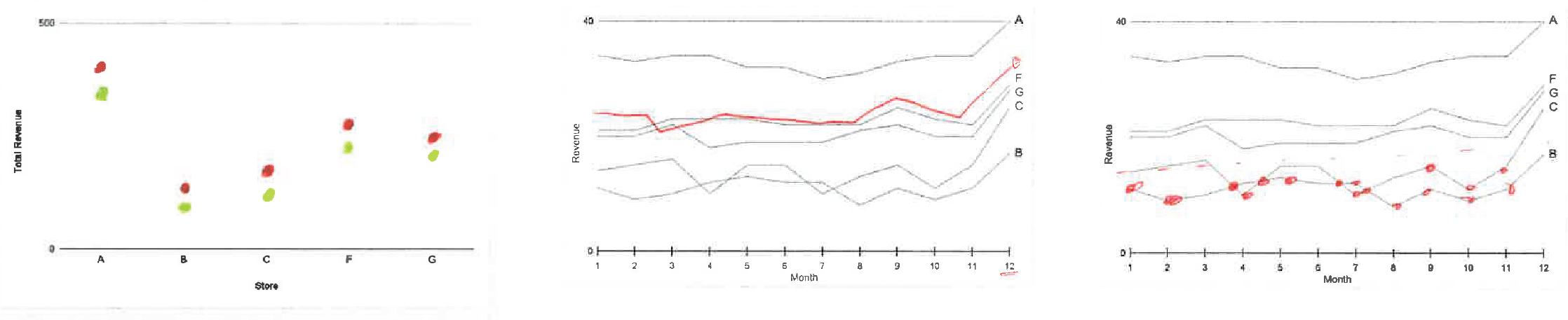}
 \caption{From left to right: examples of participants duplicating, adding, and layering marks onto existing visualizations.}
 \label{fig:add_mark_examples}
\end{figure}

\begin{figure}[H]
 \centering
 \includegraphics[width=\columnwidth]{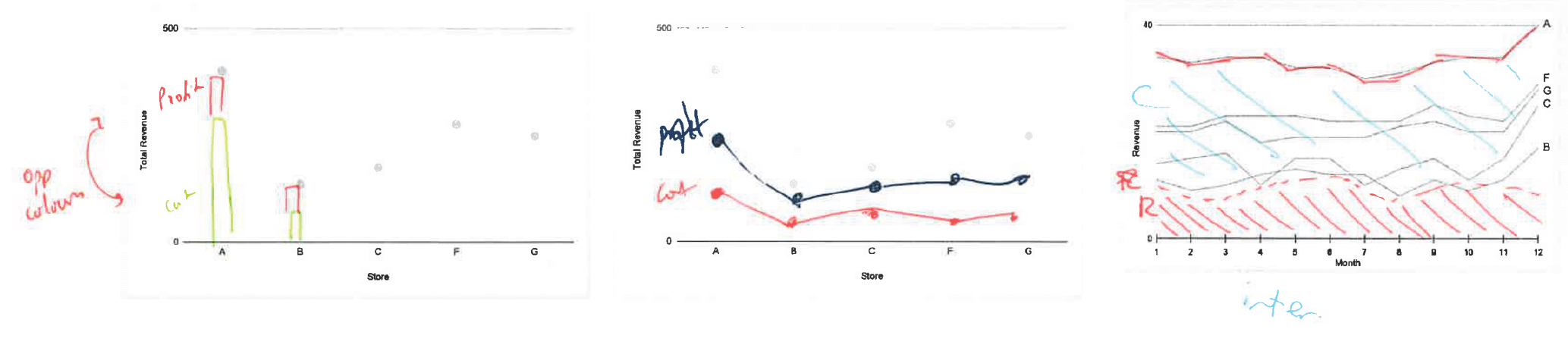}
 \caption{Examples of participants changing the mark used in a visualization. From left to right: changing a point chart to a bar chart, changing a point chart to a line chart, and changing a line chart to an area chart.}
 \label{fig:change_mark_examples}
\end{figure}

\begin{figure}[H]
 \centering
 \includegraphics[width=\columnwidth]{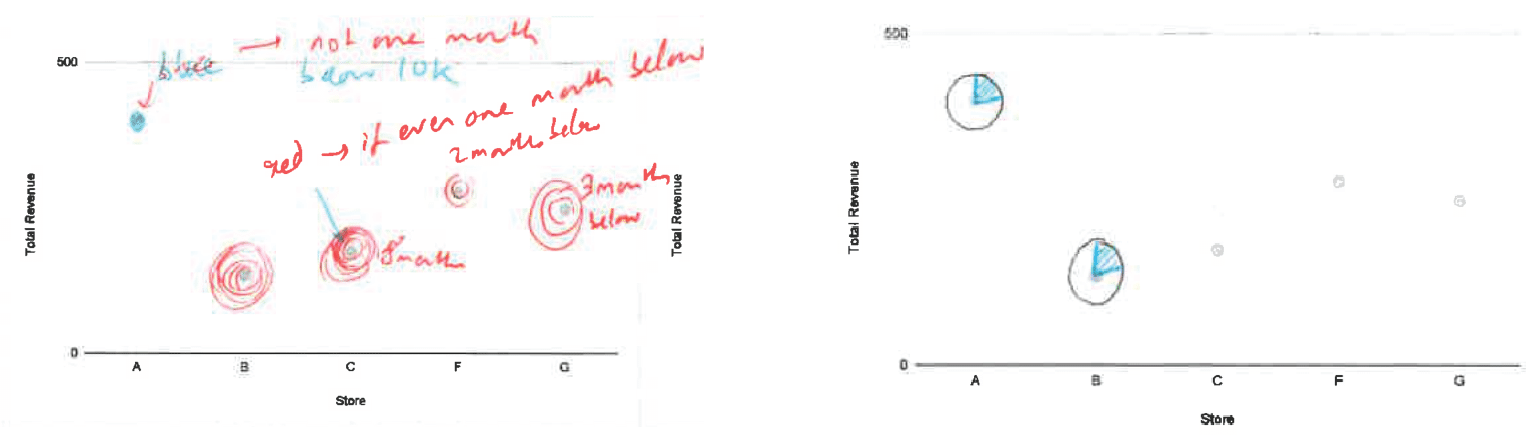}
 \caption{Two instances where participants used glyphs to convey the prompts. 
}
 \label{fig:glyph_examples}
\end{figure}

\begin{figure}[H]
 \centering
 \includegraphics[width=\columnwidth]{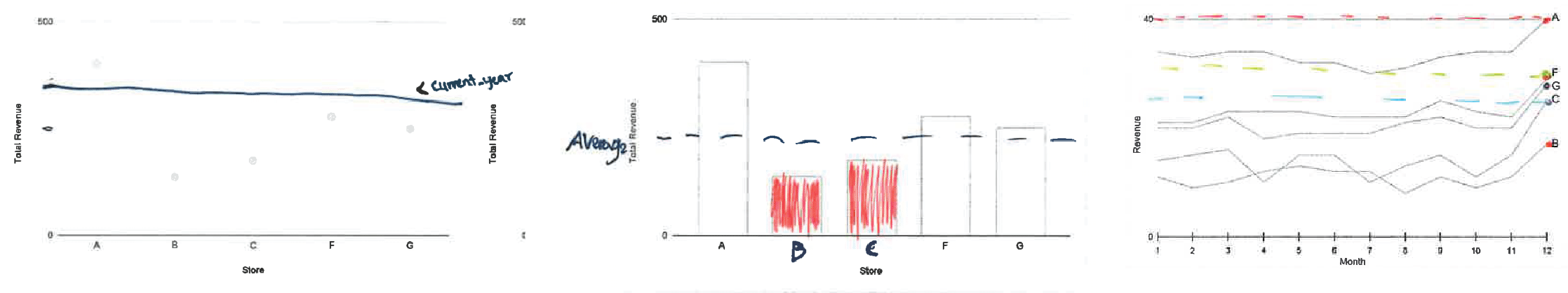}
 \caption{Examples of threshold lines added by participants. Multiple threshold lines may be added to a single visualization, and each line could be customized by style and color. 
}
 \label{fig:threshold_line_examples}
\end{figure}

\begin{figure}[H]
 \centering
 \includegraphics[width=\columnwidth]{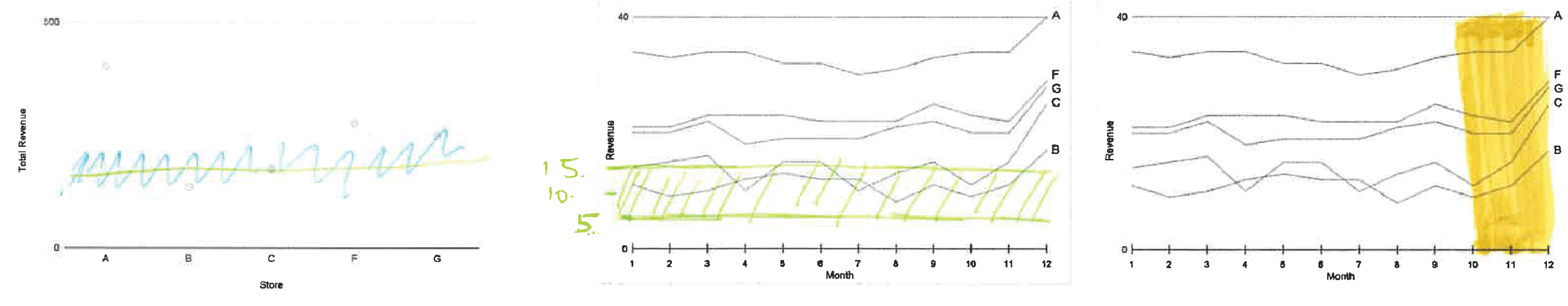}
 \caption{The visualization may be shaded to highlight important value ranges along an axis. 
}
 \label{fig:shaded_range}
\end{figure}

\begin{figure}[H]
 \centering
 \includegraphics[width=\columnwidth]{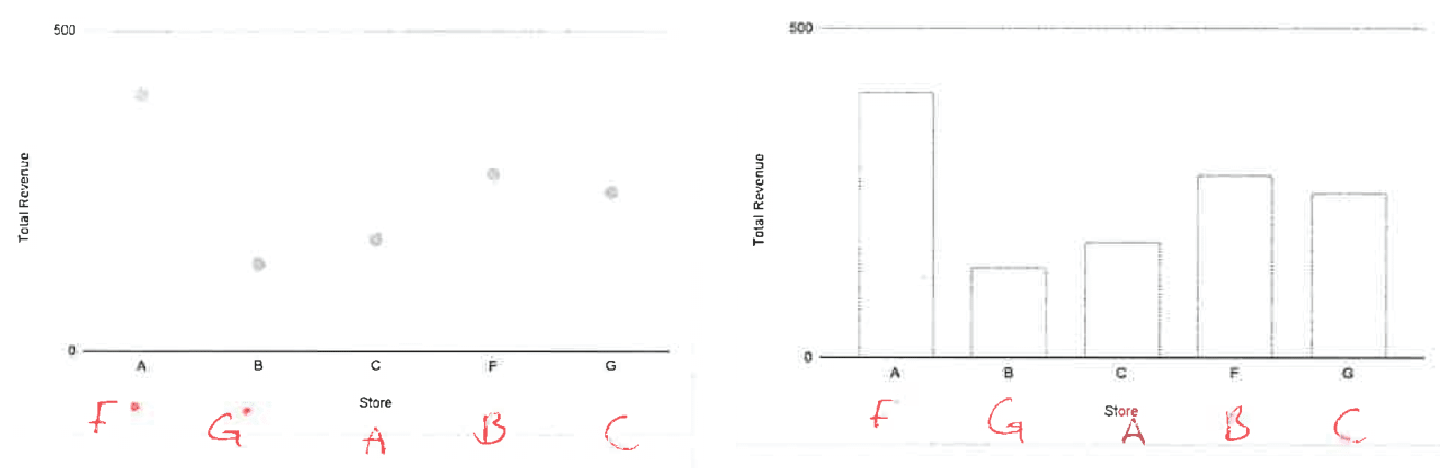}
 \caption{Two augmentations by P11 where they changed the order of elements (new order listed below) to convey priority.
}
 \label{fig:order_examples}
\end{figure}

\begin{figure}[H]
 \centering
 \includegraphics[width=\columnwidth]{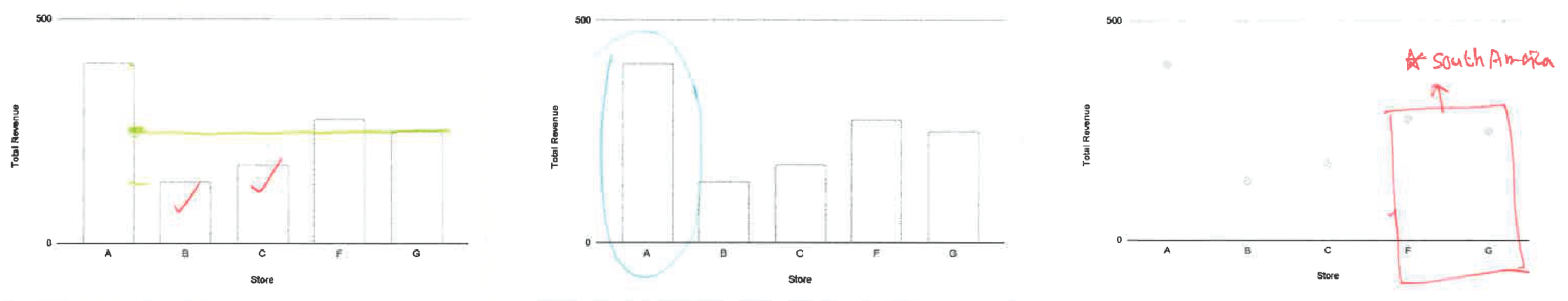}
 \caption{Examples of informal free form graphical marks made by participants.
}
 \label{fig:free_form_examples}
\end{figure}

\end{document}